\documentclass[12pt]{amsart}
\usepackage{epsfig}
\usepackage{amssymb}
\usepackage{eucal}
\newtheorem{theorem}{Theorem}[section]

\newtheorem{prop}[theorem]{Proposition}
\theoremstyle{definition}

\numberwithin{equation}{section}

\newcommand{\weglassen}[1]{}

\renewcommand{\imath}{\mathrm{i}}

\begin{document}

\title[Wannier functions of one-gap potential]
{Wannier functions of elliptic one-gap potential}

\author{E D Belokolos }
\address{Institute of Magnetism\\
Vernadski str. 36, Kiev-142\\
Ukraine\\
e-mail:\quad bel@im.imag.kiev.ua}
\author{V Z Enolskii }
\address{Dipartimento di Scienze Fisiche ''E.R.Caianiello"\\
via S.Allende, 84081  Baronissi (SA)\\
Italy\\
e-mail:\quad vze@ma.hw.ac.uk}
\author{M Salerno }
\address{Dipartimento di Scienze Fisiche ''E.R.Caianiello"\\
via S.Allende, 84081\\Baronissi (SA)\\
Italy\\e:mail:\quad salerno@sa.infn.it}
\date{\today}

\maketitle
\begin{abstract}Wannier functions of the one dimensional Schr\"odinger
equation with elliptic one gap potential are explicitly
constructed. Properties of these functions are analytically and
numerically investigated. In particular we derive an expression
for the amplitude of the Wannier function in the origin, a power
series expansion valid in the vicinity of the origin and an
asymptotic expansion characterising the decay of the Wannier
function at large distances. Using these results we construct an
approximate analytical expression of the Wannier function which is
valid in the whole spatial domain and is in good agreement with
numerical results.
\end{abstract}

\section{Introduction}

The spectral analysis of Schr\"odinger operators with periodic
potentials has been  investigated since the arbour of quantum
mechanics. In spite of this, it still represents a non exhausted
topic of ever continuing interest. From one side, it plays a
fundamental role in condensed matter physics where it provides the
mathematical basis of the quantum theory of solids. From the
other, Schr\"odinger operators with periodic and quasi-periodic
potentials play an important role in the integration of the
Kortweg-de Vries (KdV) equation. Eigenstates of these operators,
also called Bloch functions (BF), have been extensively studied
during the past years by several authors (see
\cite{zmnp80,bbeim94,gh03}). An expression of the BF in terms of
hyperelliptic $\theta-$functions was given in Ref. \cite{im75}.
These studies were further developed in Ref. \cite{kr77} where an
algebro-geometric scheme for constructing solutions of non-linear
equations was given in terms of the Baker$-$Akhiezer function.
This function is uniquely defined on the Riemann surface
associated with the energy spectrum and its properties are natural
generalizations of the analytical properties of the BF of
finite-gap potentials.

Besides BF, another set of functions which play an equally
important role in condensed matter physics are the Wannier
functions (WF) \cite{wan37}. These functions are related to BF by
a unitary transformation and form  a complete set of localised
orthonormal functions spanning a Bloch band.  The properties of these
functions were first investigated  by Kohn in 1959 \cite{kohn59}
in a classical paper in which the asymptotic decay of Wannier
functions was characterised for the case of centro-symmetrical
one-dimensional potentials. Since then, a large amount of work has
been devoted to this topic and we mention here results only some of them.
The projection operator technique was developed for construction
of the Wannier functions and the Wannier functions were studied
in the n-dimensional lattices \cite{cloize64}, \cite{cloize64a}.
The localization problem for the Wannier functions was considered
in a 1-dimensional case \cite{hevan2001}. These functions
were utilized with success in new practical methods for the electron
energy calculation of solids (e.g. \cite{kohn73}) and in a number of
modern calculation problems such as the photonic crystal circuits
\cite{busch03}.The Wannier functions represent the ideal basis for
constructing effective Hamiltonians of quantum problems involving
spatial localizations induced by electric and magnetic fields
\cite{wan60},\cite{wilk98}, \cite{gkkm98}.

In spite of this, the properties of
these functions are still not fully understood and, except for
simplest cases, there are no models for which the analytical
expression of the WF can be explicitly given. On the other hand,
the recent results achieved in the field of completely integrable
systems open the possibility to investigate analytical properties
of the WF. Quite interestingly, WF have not been considered in the
field of finite gap potentials.

The present paper represents a first contribution in this
direction. In particular, we consider  WF of
Schr\"odinger operators with one-gap potentials and use the well
developed theory of elliptic functions to investigate their
properties with sufficient completeness. As a result we derive:
{\it i)} an exact value for the amplitude of the WF at the
localization site; {\it ii)} an asymptotic expansion
characterising the decay of the WF at large distances; {\it iii)}
a power series expansion valid in the vicinity of the localization
site. Using results {\it ii),}{\it iii),} we construct an
approximate analytical representation of the WF which is valid in
the whole spatial domain. These results are shown to be in very
good agreement with the WF obtained by means of
numerical methods.

The paper is organised as follows. In Section 2 we discuss the
basic properties of the BF for one gap
potentials. In particular we introduce basic definitions, discuss
the basic properties of BF and derive the analytical
expression of their normalization constants. Section 3 is devoted
to the study of the WF. After recalling the basic definitions we
derive the main results of the paper i.e. points {\it i)-iii)}
listed above. In Section 4 we construct an approximate analytical
expression of the WF and compare the results of our theory with
WF obtained from the basic definition using
numerical tools. Finally, in Section 5 we summarise the main
results of the paper and briefly discuss future developments.

\section{Properties of Bloch functions of one-gap potential}
\subsection{ The Schr\"odinger equation with
one-gap potential} In the paper we use standard notations and facts
of the theory of elliptic functions. In particular we use
the well known Weierstrass $\wp-, \sigma-$ and $\zeta-$functions.

The periodic elliptic one-gap potential $\mathcal{U}(x)$
considered in this paper, is expressed in terms of the Weierstrass
$\wp-$function as
\begin{align}\begin{split}
&\mathcal{U}(x)=-2\wp(u),\quad u=\imath x +\omega,\quad
x\in\mathbb{R},\\
&\mathcal{U}(x+nT)=\mathcal{U}(x),\quad n\in{\mathbb{N}},
\end{split}\label{elpot}
\end{align}
where $T=-2\imath\omega'$ is the period of the lattice (notice
that $T$ is real and ${\mathcal U}(x)=2\wp(\imath x+\omega)$ is a
smooth periodic real function). The Schr\"odinger equation
associated with potential (\ref{elpot})
\begin{align}
\partial_x^2\Psi(x;E)+(E-\mathcal{U}(x))\Psi(x;E)=0.
\label{schreq}
\end{align}
As is well known (see for example \cite{bees02} and references
therein), this equation admits eigenfunctions $\Psi(x;E)$ which
satisfy the Bloch condition
\begin{equation}\label{blochperiod}
\Psi(x-T;E)=e^{-\imath k(E)T}\Psi(x;E),
\end{equation}
where $E$ is the energy given by
\begin{align}
E=\wp(v),\quad v=\alpha+\omega',\quad
\alpha\in\mathbb{R},\label{energy}
\end{align}
and $k(E)$ is the quasi-momentum given by
\begin{align}
k(v)&=\zeta(v)-\frac{\eta'}{\omega'}v,\quad v=\alpha+\omega',\quad
\alpha\in\mathbb{R}.\label{qm}
\end{align}
Eigenfunctions of type (\ref{blochperiod}) are called Bloch
functions. Notice that the dependence of the quasi-momentum on the
energy (and vice versa) arises from the elimination of the
parameter $\alpha$ from Eqs. (\ref{energy}), (\ref{qm}). The BF can
be written in explicit form as
\begin{equation}
\Psi(u;v)=C(v)\frac{\sigma(v-u)}{\sigma(v)\sigma(u)} \mathrm{exp}
\{u\zeta(\alpha)\},\label{elbloch}
\end{equation}
or, alternatively, as
\begin{equation}
\Psi(u,v)=D(v)\sqrt{\wp(u)-\wp(v)} \mathrm{exp}
\left\{\frac{\wp'(v)}{2}\int^x\frac{\mathrm{d} u}{\wp(u)-\wp(v)}
\right\}, \label{elbloch1}
\end{equation}
where $C(v), D(v)$ are proper normalization constants. In the
following we shall use both representations for the BF. Notice
that the BF considered as function of $k$ instead of $\alpha$, is
periodic in the reciprocal space, with period
\[ 2\widetilde{\omega}= \frac{\imath\pi}{\omega'}.\]
BF has the following following periodicity
properties
 \begin{align}
&\Psi(u+2n\omega;v)  =\exp\{ 2n\omega
k(v)\}\Psi(u;v),\quad
k(v)=\zeta(v)-v\frac{\eta} {\omega}
,\label{quasim1}\\
&\Psi(u+2n'\omega';v)=\exp\{ 2n'\omega'
k'(v)\}\Psi(u;v), \quad
k'(v)=\zeta(v)-v\frac{\eta'} {\omega'}
 .\label{quasim2}
\end{align}
\subsection{Normalization of the Bloch function of one-gap
potentials}  Since the normalization of the BF plays an important
role in the construction of the WF (see next section), we shall
show how to compute the normalization constant, although this question
has been considered in Chapt. VIII of \cite{bbeim94}.

We normalise the  BF according to
\begin{equation}\label{normaliz}
2\pi\langle|\Psi(x,E)|^2 \rangle=1,
\end{equation} where
\begin{equation*}\langle
f(x)\rangle=\lim_{L\to\infty}\frac{1}{L}
\int_{-\frac12L}^{\frac12 L}f(x)\mathrm{d}
x.\end{equation*} The following proposition is valid.
\begin{prop}\label{prop1}
Normalised Bloch functions of elliptic one-gap potentials are of
the form
\begin{align}
\Psi(x;\alpha)=-\frac{i}{(2\pi)^{1/2}}\left[-\wp(v)-\frac{\eta'}{\omega'}\right]^{-1/2}
\frac{\sigma(v-u)}{\sigma(v) \sigma(u)}
\exp\{v\eta+(u-\omega)\zeta(v)\},\label{bf1}
\end{align}
where $u=\imath x +\omega,\quad x\in\mathbb{R};\quad
v=\alpha+\omega',\quad\alpha\in\mathbb{R}.$
\end{prop}
\begin{proof}Let us denote the normalised BF as
\begin{equation*}
\Psi(u;v)=C(v)\Phi(u;v),
\end{equation*}
where $\Phi(u;v)$ is the non-normalised BF
\begin{equation*}\Phi(u;v)=\frac{\sigma(v-u)}{\sigma(v)\sigma(u)}
e^{(u-\omega)\zeta(v)},
\end{equation*}
and $C(v)$ is the normalization constant defined by Eq.
(\ref{normaliz})
\begin{equation*}2\pi\frac{\Big|C(v)\Big|^2}{2 \omega'} \int
\limits_{\omega}^{\omega+2\omega'}
\Big|\Phi(u;v)\Big|^2\mathrm{d}u=1. \label{normper}
\end{equation*}
The complex conjugated (non-normalised) BF is
\begin{align*}\label{compBF}&\overline{\Phi}(u;v)=\frac{\overline{\sigma}(v-u)}
{\overline{\sigma}(v)\overline{\sigma}(u)}
e^{-(u-\omega)\overline{\zeta}(v)}=\frac{\sigma(\overline{v}-\overline{u})}
{\sigma(\overline{v})\sigma(\overline{u})}
e^{-(u-\omega)\zeta(\overline{v})}\\& =-
\frac{\sigma(v-2\omega'+u-2\omega)}
{\sigma(v-2\omega')}{\sigma(u-2\omega)}e^{(u-\omega)\zeta(v-2\omega')}\\&=
\frac{\sigma(v+u)}{\sigma(v)\sigma(u)}e^{-(v-\omega')2\eta-
(u-\omega)\zeta(v)},
\end{align*}
where the following elementary equalities were used
\begin{align*}&\overline{\sigma}(z)=\sigma(\overline{z}),\quad
\overline{\zeta}(z)=\zeta(\overline{z}),\quad\overline{(u-\omega)}=
-(u-\omega), \\&\overline{u}=-u+2\omega,\quad \overline{v}=
v-2\omega', \\&  \zeta(v-2\omega')=\zeta(v)-2\eta', \\&
\sigma(v-2\omega')=-\sigma(v)\exp(-(v-\omega')2\eta'),\\&
\sigma(u-2\omega)=-\sigma(u)\exp(-(u-\omega)2\eta).
\end{align*}
By multiplying the above expressions of $\Phi(u;v)$ and $
\overline{\Phi(u;v)}$ we get
\begin{align*}\Big|\Phi(u;v)\Big|^2=\frac{\sigma(v-u)\sigma(v+u)}
{\sigma^2(v)\sigma^2(u)}e^{-(v-\omega')2\eta}= [\wp(u)-\wp(v)]
e^{(v-\omega')2\eta},
\end{align*}
where in the last step we have used the well known formula
\begin{equation*}\frac{\sigma(v-u)\sigma(v+u)}
{\sigma^2(v)\sigma^2(u)}= \wp(u)-\wp(v) .
\end{equation*}
The normalization condition can be then written in the form
\begin{align*}&1=
2\pi\frac{\Big|C(v)\Big|^2}{2 \omega'} \int
\limits_{\omega}^{\omega+2\omega'}[\wp(u)-\wp(v)]
e^{(v-\omega')2\eta}
\mathrm{d}u\\&=2\pi\Big|C(v)\Big|^2\left[-\frac{\eta'}{\omega'}-\wp(v)\right]
e^{-(v-\omega')2\eta},
\end{align*}
since
\begin{equation*}\frac{1}{2 \omega'} \int
\limits_{\omega}^{\omega+2\omega'}\wp(u)\mathrm{d}u=\frac{1}{2
\omega'}[\zeta(\omega)-\zeta(\omega+2\omega')]=
-\frac{\eta'}{\omega'}.
\end{equation*}
Thus we have obtained for the normalization constant the
expression
\begin{equation}\label{normcoeff}
C(v)=\frac{e^{\imath\theta}}{(2\pi)^{1/2}}
\left[-\frac{\eta'}{\omega'}-\wp(v)\right]^{-1/2}
e^{(v-\omega')\eta},
\end{equation}
with an arbitrary phase factor
$\exp(\imath\theta),\;\theta\in\mathbb{R}.$ In the following we
fix this factor as
\begin{equation*}\exp(\imath\theta)=\exp(\omega'\eta-\imath(\pi/2)).
\end{equation*}
\end{proof}

The normalised BF, $\Psi(u;v)$, satisfy a number of useful
properties under the action of symmetry operations. For
centro-symmetrical potentials the transformation $x\to -x$ of the
lattice corresponds to a transformation in the Jacobian
$u\to\hat{u}=-u+2\omega,$ and the following propositions can be
proved.

\begin{prop}
\begin{equation*}\Psi(\hat{u};v)=\Psi(-u+2\omega;v)=\overline{\Psi}(u;v).
\end{equation*}
\end{prop}

\begin{proof}
\begin{align*}&\Psi(\hat{u};v)=\Psi(-u+2\omega;v)\\&=
\imath(2\pi)^{-1/2}\left[-\wp(v)-\frac{\eta'}{\omega'}\right]^{-1/2}
\frac{\sigma(v+u-2\omega)}{\sigma(v)\sigma(u-2\omega)}
\exp[v\eta-(u-\omega)\zeta(v)]\\&=
\imath(2\pi)^{-1/2}\left[-\wp(v)-\frac{\eta'}{\omega'}\right]^{-1/2}
\frac{\sigma(v+u)}{\sigma(v)\sigma(u)}\exp[-v\eta-(u-\omega)\zeta(v)]
=\overline{\Psi}(u;v).
\end{align*}
\end{proof}
Similarly, that the transformation $\alpha\to -\alpha$ corresponds
to $v\to\hat{v}=-v+2\omega',$ and the following proposition is
valid.
\begin{prop}
\begin{equation*}\Psi(u;\hat{v})=\Psi(u;-v+2\omega')=\overline{\Psi}(u;v).
\end{equation*}
\end{prop}

\begin{proof}
\begin{align*}&\Psi(u;\hat{v})=\Psi(u;-v+2\omega')\\&=
-\imath(2\pi)^{-1/2}\left[-\wp(v)-\frac{\eta'}{\omega'}\right]^{-1/2}
\frac{\sigma(-v-u+2\omega')}{\sigma(-v+2\omega')\sigma(u)}\\
  &\qquad\qquad\times \exp[(-v+2\omega')
\eta+(u-\omega)\zeta(v-2\omega')]\\&=
-\imath(2\pi)^{-1/2}\left[-\wp(v)-\frac{\eta'}{\omega'}\right]^{-1/2}
\frac{\sigma(v+u)}{\sigma(v)\sigma(u)}\exp[-v\eta-(u-\omega)\zeta(v)]\\
&=\overline{\Psi}(u;v).
\end{align*}
\end{proof}

The above propositions can be used to study the elementary
properties of the BF, $\Psi(x;k)\equiv\Psi(u(x);v(k))$, where
$u(x)=\imath x +\omega$ and $v(k)$ is the inverse of the function
$k(v)=\zeta(v)-(\eta'/\omega')v$. To this regard note that
\begin{equation*}
x(\hat{u})=-x(u),\quad k(\hat{v})=-k(v).
\end{equation*}
The following two properties are easily proved.

\noindent{\bf Property 1.} $\Psi(-x;k)=\overline{\Psi}(x;k),$
\begin{proof}
\begin{equation*}\Psi(-x;k)=\Psi(\hat{u},v)=
\overline{\Psi}(u;v)=\overline{\Psi}(x;k);
\end{equation*}
\end{proof}

\noindent {\bf Property 2.} $\Psi(x;-k)=\overline{\Psi}(x;k),$
\begin{proof}
\begin{equation*}\Psi(x;-k)=\Psi(u;\hat{v})=\overline{\Psi}(u;v)
=\overline{\Psi}(x;k).
\end{equation*}
\end{proof}

\section{Analytical properties of the Wannier function of elliptic one-gap potentials}
\subsection{Definition and basic properties}

In 1937 G. Wannier introduced a complete set of functions for an
electron in a lattice structure \cite{wan37}. The Wannier
functions, $W_n(x),$ are defined as

\begin{equation}\label{wannier}
W_n(x)=\left(\frac{T}{2\pi}\right)^{1/2}\int_{-\pi/T}^{\pi/T}\Psi_n(x;k)
\mathrm{d} k.\end{equation} where the integral is made on the
Brillouin zone. WF for the Schr\"odinger operator with periodic
potential $U(x), U(x-mT), m\in{\mathbb Z}$ are localised linear
combinations of all the Bloch eigenstates of a given $n-$th
spectral band. One can easily prove that if the BF is normalised
according to Eq. (\ref{normaliz}), then the WF is normalised on
the full line,
\begin{equation*}
\int_{-\infty}^{\infty}|W_n(x)|^2\mathrm{d} x=1.
\end{equation*}
Using the translation operator one then  constructs a countable
set of WF:  $W_n^{(l)}(x):=W_n(x-lT),\quad l\in{\mathbb {Z}}$
which is complete and forms an ortho-normal basis
\begin{equation*}
\int_{-\infty}^{\infty}\overline{W}_n^{(l)}(x)W_{n'}^{(l')}(x)\mathrm{d}
x=\delta_{nn'}\delta_{ll'},\quad  l\in{\mathbb {Z}}.
\end{equation*}
The inverse transformation allows to express a BF in terms of WF
as
\begin{equation}\Psi_n(x;k)=\left(\frac{T}{2\pi}\right)^{1/2}
\sum_{l=-\infty}^{\infty}W_n^{(l)}(x)e^{ilak}.
\end{equation}
In the following we shall omit the band index $n$ since we deal
only with one band. Properties of WF of one dimensional periodic
potentials  were studied by W. Kohn \cite{kohn59} where he proved
that for every band there exists one and only one WF which
satisfies simultaneously the following three properties
$1)\overline{W}(x)=W(x);\quad  2) W(-x)=\pm W(x);\quad 3)
W(x)=O(\exp(-h|x|)),$ where $h>0.$ In the following we investigate
the analytical properties of the WF for the one-gap potential in
Eq. (\ref{elpot}). In this case the WF is given by the formula

\begin{align}\label{wf24}
W(x)&=\left(\frac{T}{2\pi}\right)^{1/2}\int_{-\pi/T}^{\pi/T}\Psi(x;k)
\mathrm{d}
k \nonumber \\
&=\left(\frac{T}{2\pi}\right)^{1/2}\left(\int_{-\pi/T}^0
+\int_0^{\pi/T} \right)\Psi(x;k)\mathrm{d}
k\nonumber\\
&=\left(\frac{T}{2\pi}\right)^{\frac12} \int_0^{\pi/T}
\left(\Psi(x;k)+\Psi(x;-k)\right)\mathrm{d} k\nonumber\\
&= \left(\frac{T}{2\pi}\right)^{1/2} 2 \mathrm{Re}\int_0^{\pi/T}
\Psi(x;k)\mathrm{d} k\nonumber\\
&= \mathrm{Re}\left\{ -\imath\frac{\sqrt{-2\imath\omega'}}{\pi}
\int_{\omega'}^{\omega+\omega'}\sqrt{\frac{\mathrm{d}k(v) }{\mathrm{d}v}    }\frac{\sigma(v-u)}
{\sigma(v)\sigma(u)}
e^{v\eta+(u-\omega)\zeta(v)}\mathrm{d}v\right\}.
\end{align}
Using the properties of the BF, $\Psi(x;k)$, the following basic
properties of the WF can be proved.
\begin{prop} $\overline{W}(x)=W(x)$. \end{prop}
\begin{proof} \begin{align*}
\overline{W}(x)&=\left(\frac{2T}{\pi}\right)^{1/2}
\mathrm{Re}\int_{0}^{\pi/T} \overline{\Psi}(x;k)\mathrm{d}k
\\
&=\left(\frac{2T}{\pi}\right)^{1/2}
\mathrm{Re}\int_{0}^{\pi/T}\Psi(x;k)\mathrm{d} k =W(x).\end{align*}
\end{proof}
\begin{prop} $W(-x)=W(x)$.
\end{prop}
\begin{proof}\begin{align*}
&W(-x)=\left(\frac{2T}{\pi}\right)^{1/2}
\mathrm{Re}\int_{0}^{\pi/T}\Psi(-x;k)\mathrm{d}k=
\left(\frac{2T}{\pi}\right)^{1/2} \mathrm{Re}\int_{0}^{\pi/T}
\overline{\Psi}(x;k)\mathrm{d}k\\& =\left(\frac{2T}{\pi}\right)^{1/2}
\mathrm{Re}\int_{0}^{\pi/T} \Psi(x;k)\mathrm{d}k=W(x).
\end{align*}
\end{proof}

\subsection{Power series expansion of the Wannier function at x=0}
We shall construct in this section the power series expansion
of the Wannier function of one gap potential.
\begin{theorem}
The Wannier function of the lower energy band for the one gap potential
admits the following power
series representation
\begin{equation}\label{WFpowexp}
W(x)=\sum_{p=0}^{\infty}\frac{(-1)^p}{(2p)!}W_{2p}x^{2p},
\end{equation}
where the coefficients $W_{2p}$ of the expansion (\ref{WFpowexp})
are given by the formula
\begin{align}W_{2p}=\sum_{l=0}^{p}M_{l}q_{p,l}.
\end{align}
Here
\begin{align}
M_l&=\frac{\sqrt{2\imath}}{\pi}\sqrt{\omega'e_3+\eta'}\sum_{j=0}^l
\frac{(2j-1)!!l!}{2^j (j!)^2(l-j)!}e_3^j(e_2-e_3)^{l-j}\label{moments}\\
&\times F\left(-\frac12,j+\frac{1}{2};j+1;
\frac{\omega'(e_3-e_2)}{\omega'e_3+\eta'}  \right) \nonumber ,
\end{align}
where $F(a,b;c;z)$ is the standard
hypergeometric function, $e_2,e_3$ are branch points of the elliptic
curve and $q_{p,l}$ are coefficients of polynomials in $\wp(v)$
\[Q_{p}(\wp(v))=\sum_{l=0}^pq_{p,l}\wp^{l}(v)  \]
defined by the recurrence
\begin{equation}
Q_{p}(\wp(v))=\sum_{m=0}^{p-1}\left(\begin{array}{c}
2p\\2m-2\end{array}\right)\phi_{m-p-1}Q_m(\wp(v))
\end{equation}
with
\[\phi_0=2e_1+\wp(v),\quad \phi_{p}=2\wp^{(2p)}(\omega).
\]

First few coefficients of the expansion (\ref{WFpowexp}) are
\begin{align}
\begin{split}
W_0&=M_0,\\
W_2&=M_1+2e_1  M_0, \\
W_4&=M_2+4 e_1M_1+(4e_1^2
+2\wp''(\omega))M_0,\\
W_6&=M_3+6e_1M_2+(14\wp''(\omega)+12e_1^2)M_1\\
&+(2\wp^{(IV)}(\omega)+28\wp''(\omega)e_1+8e_1^3) M_0.
\end{split}
\label{firstw0}
\end{align}
\end{theorem}

\begin{proof}
We have
\begin{equation}
W(x)=\left( \frac{2T}{\pi}  \right)^{\frac12}\mathrm{Re}
\int\limits_{0}^{\pi/T}\Psi(x;k)\mathrm{d}k.\label{wannier1}
\end{equation}
Because the BF $\Psi(x;k)$ is even
 in $x$ we can write it in the form of Taylor expansion
\begin{equation}\Psi(u,v)=\sum_{p=0}^{\infty}
\frac{(-1)^p}{(2p)!}\Psi_{2p}(v)x^{2p},
 \label{taylor1}\end{equation}
where
\[ \Psi_{2p}(v)=\left[\frac{\mathrm{d}^{2p} }{\mathrm{d} u^{2p} }\Psi(u,v)
\right]_{u=\omega}, \quad p=1,\ldots.
 \]
If we substitute the Taylor expansion (\ref{taylor1}) to the
(\ref{wannier1}) we obtain the expansion (\ref{WFpowexp}) with the
following coefficients
\begin{equation}\label{expansion}
W_{2p}= \left(\frac{2 T}{\pi}\right)^{1/2}
\mathrm{Re}\int\limits_0^{\pi/T}
\Psi_{2p}(v)\mathrm{d} k.
\end{equation}

Using Schr\"odinger equation we obtain easily for the $\Psi_{2p}(v)$ a
recurrent relation
\begin{align*}
\Psi_{2p}(v)&=\sum_{l=0}^{p-1}\left(\begin{array}{c}
2p\\2l-2\end{array}\right)\phi_{p-l-1}\Psi_{2l}(v)\\
\phi_0&=2e_1+\wp(v),\quad \phi_{p}=2\wp^{(2p)}(\omega).
\end{align*}
The form of this relation leads to conclusion that
\[ \Psi_{2p}(v)=Q_p(\wp(v))\Psi_0(v), \]
where
\begin{align}\label{coeffpsi}
\Psi_0(v)&=\Psi(\omega;v)=\frac{1}{\sqrt{2\pi}}
\sqrt{\frac{\wp(v)-e_1} {\wp(v)+\frac{\eta'}{\omega'} }}
\end{align}
and $Q_p(\wp(v))$ are polynomials of the $p$-th order in $\wp(v)$,
\[Q_p(\wp(v))=\sum_{l=0}^p q_{p,l}\wp^l(v).\]
Similarly to $\Psi_{2p}(v)$ the polynomials $Q_{p}(\wp(v))$ satisfy the
following recurrent relation
\begin{equation}
Q_{p}(\wp(v))=\sum_{m=0}^{p-1}\left(\begin{array}{c}
2p\\2m-2\end{array}\right)\phi_{m-p-1}Q_m(\wp(v))
\end{equation}
with
\[\phi_0=2e_1+\wp(v),\quad \phi_{p}=2\wp^{(2p)}(\omega)
.\]

In particular the first few polynomials
$Q_p(\wp(v))$ are
\begin{align}
\begin{split}
Q_1(\wp(v))&=\wp(v)+2e_1,\\
Q_2(\wp(v))&=\wp(v)^2+4e_1\wp(v)+
2\wp''(\omega)+4e_1^2,\\
Q_3(\wp(v))&=\wp(v)^3+6e_1\wp(v)^2+
\left(14\wp''(\omega)+12e_1^2\right)\wp(v)\\&
+2\wp^{(IV)}(\omega)+28e_1\wp''(\omega)+8e_1^3.
\end{split}
\label{blochrec}
\end{align}

Next we calculate the integral expressions of the coefficients
$W_{2p}.$  We show that the following formula is valid
\begin{align*}
M_l&=-2\left(\frac{\omega'}{\imath\pi}\right)^{1/2}\int\limits_{0}^{\omega'}
\wp(v)^l\left(\wp(v)+\frac{\eta'}{\omega'}\right)\Psi(\omega;v)\mathrm{d}v
\\
&=\left(\frac{2T}{\pi}\right)^{\frac12}\int\limits_{0}^{\omega'}\wp(v)^l
\frac{1}{\sqrt{2\pi}}\sqrt{\frac{\wp(v)-e_1}{\wp(v)+\frac{\eta'}{\omega'}}}
\left(-\wp(v)-\frac{\eta'}{\omega'}\right)\mathrm{d}v
\\
&=\frac{T^{\frac12}}{\pi}\int\limits_{0}^{\omega'}\wp(v)^l
\sqrt{\wp(v)-e_1}\sqrt{\wp(v)+\frac{\eta'}{\omega'}}\mathrm{d}v.
\end{align*}
After the substitution $\wp(v)=s$, the computation is reduced to
the derivation of the complete elliptic integral
\begin{equation}
M_l=-\frac{\sqrt{-\imath\omega'}}{\pi}\int\limits_{e_3}^{e_2}
s^l\sqrt{\frac{s+\frac{\eta'}{\omega'}}{(s-e_2)(s-e_3)}}\mathrm{d}
s.\label{integral1}\end{equation}
By introducing the new variable
\[t= \frac{s-e_3}{e_2-e_3}, \]
the integral $M_l$ acquires the form
\begin{equation}
M_l=-\frac{\sqrt{-\imath\omega'}}{\pi}\sqrt{e_2-e_3}\int\limits_{0}^{1}
((e_2-e_3)t-e_3)^l\sqrt{\frac{1-\widetilde{k}^2t}{t(1-t)}}\mathrm{d}
 t,\label{integral}\end{equation}
where
\begin{equation}  \widetilde{k}=\sqrt{\frac{\omega'(e_3-e_2)}
{\omega'e_3+\eta'}},\label{auxilmod} \end{equation}
is the Jacobi modulus of the elliptic curve
\begin{equation}
Y^2=(X-e_2)(X-e_3)\left(X+\frac{\eta'}{\omega'}\right).
\end{equation}

Using the integral representation of the hypergeometric
function (see e.g. \cite{as65})
\begin{equation}F(a,b;c;z)=\frac{\Gamma(c)}{\Gamma(b)\Gamma(c-b)}
\int_0^1\frac{t^{b-1}(1-t)^{c-b-1}}{(1-tz)^{a}}\mathrm{d} t,
\end{equation}
we obtain the required expression
\begin{align}\label{finalM}\begin{split}
M_l&=\frac{\sqrt{2\imath}}{\pi}\sqrt{\omega'e_3+\eta'}\sum_{j=0}^l
\frac{(2j-1)!!l!}{2^j(j!)^2 (l-j)!}e_3^j(e_2-e_3)^{l-j}\\
&\times F\left(-\frac12,j+\frac12;j+1;\widetilde{k}^2\right) .
\end{split}\end{align}
\end{proof}

It is worth to note that the coefficient $W_0$ gives an exact
value of the amplitude of WF at the localization site $x=0,$
\begin{align}
W_0=\frac{\sqrt{2\imath}}{2}\sqrt{\omega'e_3+\eta'}
F\left(-\frac12,\frac12;1;\widetilde{k}^2\right).
\label{amplitude1}
\end{align}
This amplitude can also be written in the alternative
form
\begin{align}
W_0=\frac{\sqrt{2\imath}}{\pi}\sqrt{\omega'e_3+\eta'}E(\widetilde{k}),
\label{amplitude}
\end{align}
where $E(\widetilde{k})$ is the complete integral of the second
kind depending on $\widetilde{k}$.
Also note that by using the relations
\begin{align*}
&\left( z\frac{\mathrm{d}}{\mathrm{d} z } +b  \right)= b F(a,b+1;c;z),\\
&\left[(1-z) \frac{\mathrm{d}}{\mathrm{d} z } +c-a-b
\right]F(a,b;c;z)=\frac{(c-a)(c-b)}{c}F(a,b;c+1;z),
\end{align*}
one can express all hyperegeometric functions
$F(-\frac12,j+\frac12;j+1; \widetilde{k}^2)$ in (\ref{finalM})
in terms of the derivatives of
$F(-\frac12,\frac12;1;\widetilde{k}^2)$ with respect to
$\widetilde{k}^2$, and therefore the whole expression can be written
in terms of the complete integral $E(\widetilde{k})$ and of its derivatives.

We remark that the quantities $\wp^{(2j)}(\omega), j=1,\ldots, $
can be computed in recurrent way, the some first of them are \cite{as65}
\begin{align*}
\wp''(\omega)&=3!\left(e_1^2-\frac{1}{2^2\cdot 3}g_2\right),\\
\wp^{(IV)}(\omega)&=5!\left(e_1^3-\frac{3}{2^2\cdot 5}g_2e_1
-\frac{1}{2\cdot 5}g_3\right),\\
\wp^{(VI)}(\omega)&=7!\left(e_1^4-\frac15g_2e_1^2-\frac17g_3e_1
+\frac{1}{2^4\cdot 5 \cdot 7}g_2^2  \right).
\end{align*}

\subsection{Asymptotic expansion of the Wannier function}
In this section we obtain the asymptotic expression for the WF at
$x\to +\infty$ by the steepest descent method. This method (see,
e.g. \cite{fed87})  permits to compute the asymptotic expression
of integrals of the type
\[ F(x)=\int\limits_{\gamma} f(z)\mathrm{exp}\{x S(z)\}\mathrm{d}z ,  \]
where $\gamma$ is a contour in the complex plane and the functions
$f(z)$ and $S(z)$ are holomorphic in the vicinity of $\gamma$.
When a saddle point $z_0,$ which is defined by the equation
\begin{equation*}
\frac{\mathrm{d}}{\mathrm{d}  z} S(z_0) =0,
\end{equation*}
does not coincide with the edges of the contour, the asymptotic
formula of the integral $F(x)$ reads
\[ F(x)=
\sqrt{-\frac{2\pi}{x\frac{\mathrm{d}^2S(z_0)}{\mathrm{d} z^2}}}
  \mathrm{exp}\{x S(z_0) \} \left[f(z_0) +O(x^{-1}) \right].
\]
In our case we must use a non-standard variant of the steepest
descents method, since the exponential in the integrand will have
the form $xS(z)$ only at $|x|\to +\infty$ and $f(z_0)=0$.

\begin{prop}
\label{wasympt}At $x\to\infty$ the Wannier function of the lower
energy band for the one gap potential has the
following asymptotic expression
\begin{equation} W(x)\simeq \mathrm{Re}
\left\{ \frac{\sqrt{-2\imath \omega'}}{\pi}
     \left(
e_1+\frac{\eta^{\prime}}{\omega^{\prime}}\right)^{1/2} \frac{\sigma(v
-u)}{\sigma(v)\sigma(u)}e^{(u-\omega)\zeta(v)}
 \left[ \frac{\imath}{2\wp^{\prime}(v)}\right]^{1/4}
\frac{\Gamma\left(\frac34\right)}{x^{\frac34}}\right\},\label{asform}
\end{equation}
where $v$ is a solution of the equation
\begin{equation}\wp(v)=-\frac{\eta^{\prime}}{\omega^{\prime}},
\quad\mbox{or}\quad k^{\prime}(v)=0,\label{wpzero}
\end{equation}
such that the complex number
$\omega'k(v)=\omega'(\zeta(v)-(\eta'/\omega')v)$ has negative real
part. This means that $$W(x)\simeq \exp(-h\vert x\vert)\vert
x\vert ^{-3/4},\quad \vert x\vert\to\infty,$$ where $h=|k(v)|$ and
$v$ is defined by the equation $k'(v)=0.$\end{prop}

\begin{proof} We use the expression of the WF for the first
energy band $[e_3, e_2]$ given in Eq. (\ref{wf24}).
Since
\begin{align*}&\frac{\sigma(v-u)}{\sigma(v)\sigma(u)}=-
\frac{\sigma(u-v)}{\sigma(\omega-v)}\frac{\sigma(\omega)}{\sigma(u)}
\frac{\sigma(\omega-v)}{\sigma(v)\sigma(\omega)}\\&=
\frac{\sigma(v-\omega)}{\sigma(v)\sigma(\omega)}\exp\left\{
\int_{\omega}^{u}[\zeta(s-v)-\zeta(s)] \mathrm{d}s\right\}\\&=
[\wp(v)-e_1]^{1/2}\exp\left\{-v\eta+
\int_{\omega}^{u}[\zeta(s-v)-\zeta(s)] \mathrm{d}s\right\},
\end{align*}
we  have that Eq. (\ref{wf24}) can be rewritten as
\begin{align}
\label{asyexp} W(x)&= \mathrm{Re}\left\{\frac{\sqrt{-2\imath
\omega'}}{\pi}
\int_{\omega'}^{\omega+\omega'}[e_1-\wp(v)]^{1/2}\sqrt{\frac{\mathrm{d}k(v) }{\mathrm{d}v}    }\right.\\
&\left.\times\exp\left\{\int_{\omega}^{u}[\zeta(s-v)-\zeta(s)+\zeta(v)]
\mathrm{d}s\right\} \mathrm{d}v\right\}.\nonumber
\end{align}

When $x\to +\infty$ we can calculate the integral in Eq.
(\ref{asyexp}) by the steepest descents method. To this regard we
remark that the argument of the exponential, as a function of $v$,
has saddle points which are defined by the equation
\begin{equation*}\frac{\mathrm{d}}{\mathrm{d}v}
\int_{\omega}^{u}[\zeta(s-v)-\zeta(s)+\zeta(v)]
\mathrm{d}s=0,
\end{equation*}
or, in other words, by the equation
\begin{equation*}\wp(v)=-\frac{\zeta(u-v)-\zeta(\omega-v)}{u-\omega}.
\end{equation*}
At $x\to +\infty$ the last equation attains the form
(\ref{wpzero}). Since $\overline{\wp}(v)=\wp(\overline{v})$ we
have that if $v$ is saddle point then $\overline{v}$ is also a
saddle point. On the other hand, $\wp(v)$ is an elliptic function
of the second order, so it takes every value twice in the
fundamental domain, this implying that there are two saddle
points, say  $v_1, v_2$, in  the fundamental domain. The sum of
the values $v_1, v_2$ must be a period of the lattice, which in
our case means that $v_1 +v_2=2(\omega+\omega').$ It is not
difficult  to show that
\begin{equation*}v_{1}=\omega+\omega'+\imath\beta,\quad
v_{2}=\omega+\omega'-\imath\beta, \quad\beta\in\mathbb{R},
\end{equation*}
i.e. the two saddle points $v_{1}, v_{2},$ are  situated in the
spectral gap. The periodicity of the Weierstrass function $\wp(z)$
in the complex plane give rise to countable set $V$ of saddle
points,
\begin{equation*}V=\{v_{1}+2n_1\omega', v_{2}+2n_2\omega':
n_1, n_2\in \mathbb{Z}\}.
\end{equation*}
In order to build the proper asymptotic expression for the Wannier
function $W(x)$ we must select from this set a special  saddle
point which we denote by $v_0$. In the neighbourhood of $v_0$ we
have
\begin{equation*}\sqrt{\frac{\mathrm{d}k(v) }{\mathrm{d}v}    }
\simeq [-\wp'(v_0)]^{1/2}(v-v_0)^{1/2},
\end{equation*}
\begin{align*}&\int_{\omega}^{u}[\zeta(s-v)-\zeta(s)+\zeta(v)]
\mathrm{d}s\simeq\int_{\omega}^{u}[\zeta(s-v_0)-\zeta(s)+\zeta(v_0)]
\mathrm{d}s\\&+\frac{1}{2}(v-v_0)^2\int_{\omega}^{u}[\zeta''(s-v_0)+
\zeta''(v_0)]
\mathrm{d}s=\int_{\omega}^{u}[\zeta(s-v_0)-\zeta(s)+\zeta(v_0)]
\mathrm{d}s\\&-(u-\omega)\frac{1}{2}\left[\wp'(v_0)+\frac{\wp(u-v_0)-
\wp(\omega-v_0)}{u-\omega}\right](v-v_0)^2\\&\simeq
\int_{\omega}^{u}[\zeta(s-v_0)-\zeta(s)+\zeta(v_0)]
\mathrm{d}s-(u-\omega)\frac{1}{2}\wp'(v_0)((v-v_0)^2.
\end{align*}
Substituting the last two expressions into the integral
representation of the  WF in (\ref{asyexp}) , we
obtain
\begin{align*}&W_0(x)
\simeq \mathrm{Re}\left\{ \frac{\sqrt{-2\imath\omega'}}{\pi}  \left(
e_1+\frac{\eta^{\prime}}{\omega^{\prime}}\right)^{1/2}
\exp\left\{\int_{\omega}^{u}[\zeta(s-v_0)-\zeta(s)+\zeta(v_0)]
\mathrm{d}s\right\}\right.\\&\left.\times\int_{C_0}
\mathrm{d}v[-\wp'(v_0)]^{1/2}(v-v_0)^{1/2}
 \exp\left\{-\frac{1}{2}
(u-\omega)\wp'(v_0)(v-v_0)^2\right\}\right\}\\&=
\mathrm{Re}\left\{ \frac{\sqrt{-2\imath\omega'}}{\pi}   \left(
e_1+\frac{\eta^{\prime}}{\omega^{\prime}}\right)^{1/2}
\frac{\sigma(v_0-u)}{\sigma(v_0)\sigma(u)}\exp\left\{(u-\omega)
\zeta(v_0)\right\}\right.
\\&\left.\times\int_{C_0}
\mathrm{d}r[-\wp'(v_0)]^{1/2}r^{1/2}\exp\left\{-\frac{1}{2}
(u-\omega)\wp'(v_0)r^2\right\}\right\},
\end{align*}
where $C_0$ is a contour passing through the saddle point $v_0.$
The integral in the last expression can be calculated as
\begin{align*}&I_0=\int_{C_0}
\mathrm{d}r[-\wp'(v_0)]^{1/2}r^{1/2}\exp\left\{-\frac{1}{2}
(u-\omega)\wp'(v_0)r^2\right\}\\&=\left[\frac{\imath}{2\wp'(v_0)}\right]^{1/4}
\frac{1}{x^{\frac34}}\int_{0}^{\infty}e^{-t}t^{-1/4}\mathrm{d}t=
\left[\frac{\imath}{2\wp'(v_0)}\right]^{1/4} \frac{\Gamma\left(\frac34
\right)}{x^{\frac34}}.
\end{align*}
Notice that, since $\wp^{\prime}(v_0)=-2[\wp(v_0)-e_1]^{1/2}
[\wp(v_0)-e_2]^{1/2}[\wp(v_0)-e_3]^{1/2},$  and $e_3\leq
e_2\leq\wp(v_0)\leq e_1,$ we have that
$\wp^{\prime}(v_0)=-\imath|\wp^{\prime}(v_0)|.$

For the function $W_0(x)$ we finally obtain
\begin{align*}
&W_0(x)\simeq \mathrm{Re}\left\{\frac{\sqrt{-2\imath \omega'} }{\pi}
\left( e_1+\frac{\eta^{\prime}}{\omega^{\prime}}\right)^{1/2}
\left[\frac{\imath}{2\wp'(v_0)}\right]^{1/4}
\frac{\Gamma\left(\frac34\right)}{x^{\frac34}}\right.\\
&\left.\times\exp\left\{\int_{\omega}^{u}[\zeta(s-v_0)-\zeta(s)+\zeta(v_0)]
\mathrm{d}s\right\}\right\}\\&=
\mathrm{Re}\left\{\frac{\sqrt{-2\imath \omega'} }{\pi} \left(
e_1+\frac{\eta^{\prime}}{\omega^{\prime}}\right)^{1/2}
\frac{\sigma(v_0 -u)}{\sigma(v_0)\sigma(u)}e^{(u-\omega)\zeta(v_0)}
 \left[ \frac{\imath}{2\wp^{\prime}(v_0)}\right]^{1/4}
\frac{\Gamma\left(\frac34\right)}{x^{\frac34}}\right\}.
\end{align*}
The asymptotic behaviour of the function $W_0(x)$ at
$x\to+\infty$ is defined by the factor
\begin{equation*}f(u,v_0)=\frac{\sigma(v_0-u)}{\sigma(u)}\exp[(u-
\omega)\zeta(v_0)],
\end{equation*}
which satisfies the relation
\begin{equation*}f(u+2n\omega',v_0)=f(u,v_0)\exp[2n\omega'(\zeta(v_0)
-\frac{\eta'}{\omega'}v_0)]=f(u,v_0)\exp[2n\omega'k(v_0)].
\end{equation*}
The saddle point $v_0$ must be chosen
in such a manner that the complex number $\omega'k(v_0)$ has
negative real part. From the  previous considerations it follows that
the  point $\overline{v}_0$ is also a saddle point. The asymptotic
behaviour of the function $W_0(x)$ at $x\to -\infty$ is defined by
this saddle point which corresponds to the complex number
$\omega'k(\overline{v}_0)=\omega'(\zeta(\overline{v}_0)-(\eta'/\omega')
\overline{v}_0)$ with positive real part.
\end{proof}

It is appropriate to make here some remarks.

According to the theorem $$W(x)\simeq\exp(-|x||k(\alpha_0)|)\vert x
\vert^{-3/4},
\quad \vert x\vert\to\infty.$$
It is easy to understand such an asymptotic behaviour of the WF
at $|x|\to\infty$ if we take into account that
$$W(x)\simeq \mathrm{Re}\left\{\int_{C}(k-k_0)^{\beta}
\mathrm{e}^{\imath kx}\right\}
\simeq 2\sin(\beta\pi)\Gamma(1+\beta)x^{-(1+\beta)}e^{-\Im k_0 x},$$
where $k_0$ is a branching point of the energy $E(k),$ and that, due to
a normalization constant of the wave function,
$$\Psi(k)\simeq(k-k_0)^{-1/4},$$ the equality $\beta=-1/4$ is
valid. As far as we know
this asymptotic law was mentioned for the first time in
\cite{hevan2001}.

It is of interest to note also that the equation
\[\quad \wp(v)+\frac{\eta'}{\omega'}=0  \]
has obviously the following solution
\[v=\pm\int\limits_{-\frac{\eta'}{\omega'}}^{\infty}
\frac{\mathrm{d} x}{\sqrt{4 x^3-g_2 x - g_3}}.\]
More general problem to solve the equation
\[ \wp(v,\omega,\omega')=c(\omega,\omega'), \]
is a well known mathematical problem in the theory of elliptic functions.
Solution of the problem in terms of Eisenstein series
is presented in the paper\cite{ez82}.

In the above theorem we have obtained results for the Wannier function of
lower energy band. Results for the Wannier function of higher band,
\begin{align}\begin{split}
W(x)&=\mathrm{Re} \left\{
-\imath\frac{\sqrt{-2\imath\omega'}}{\pi}
\int_{\omega}^{\widetilde{\omega}} \sqrt{
\frac{\mathrm{d}k(v)}{\mathrm{d}v}
}\frac{\sigma(v-u)}{\sigma(v)\sigma(u)}
\mathrm{e}^{v\eta+(u-\omega)\zeta(v)}\mathrm{d}v\right\},\\
 k(\widetilde{\omega})&=\zeta(\widetilde{\omega})-
\frac{\eta'}{\omega'}\widetilde{\omega}=
\frac{\pi}{T},  \end{split}  \label{wf24a}
\end{align}
are similar to ones presented above and as a result of that we
omit appropriate considerations.

Let us now discuss two limiting cases. The limit
$\tau=\omega'/\omega\to 0$ corresponds to the free electron case
or the empty lattice case. In this limit the energy gap is zero,
$e_1=e_2,$ there are no saddle points and as a result we have the
well known free electron WF
$$W(x)=\frac{T^{1/2}}{\pi x} \sin\left(\frac{\pi x}{T}\right).$$
The limit $\tau=\omega'/\omega\to \imath\infty$ corresponds to the
case of tightly bound electrons. In this case the width of the
lower energy band is zero, $e_2=e_3,$ the appropriate wave
function looks as follows,
$$\Psi(x)=\left(\frac{\alpha}{2}\right)^{1/2}\frac{1}{\cosh\alpha x},\quad
E_0=-\alpha^2,$$
where $E_0$ is the binding energy.
The wave functions of higher energy bands are of a form,
$$\Psi(x,k)=\frac{1}{\sqrt{2\pi}\sqrt{k^2+\alpha^2}}
\left(|k|+\imath\alpha\tanh(\alpha x)
\right)e^{\pm \imath k x},\quad E=k^2.$$

In the next section we shall
compare our analytical results with numerical ones.
\begin{figure}
\centering \resizebox{0.65\textwidth}{!}{
\includegraphics[clip]{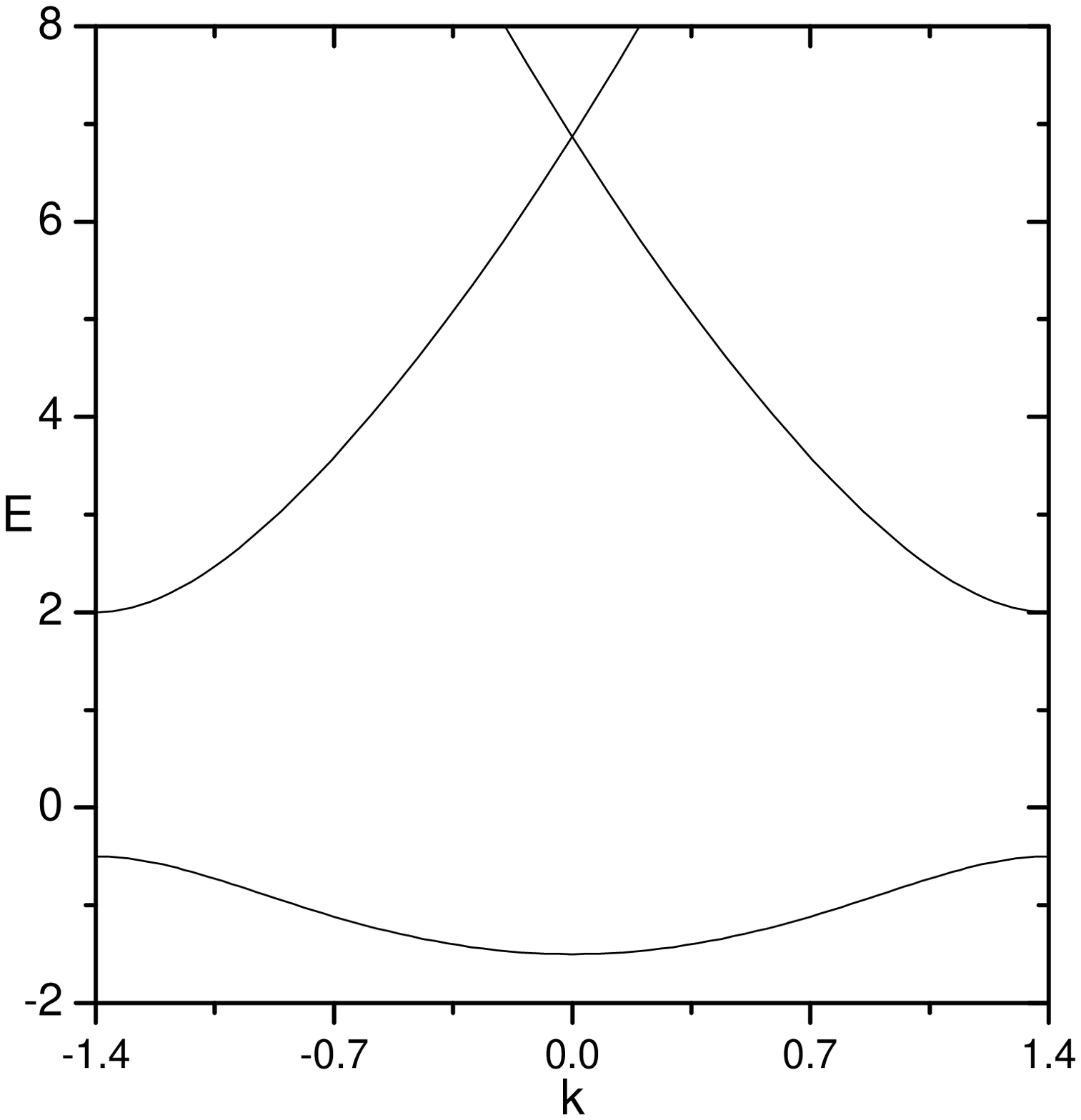}
} \caption{ Energy bands of the one gap potential with parameters
of the elliptic curve $e_1=2,e_2=-0.5,e_3=-1.5$} \label{fig-1}
\end{figure}
\begin{figure}
\centering \resizebox{0.65\textwidth}{!}{
\includegraphics[clip]{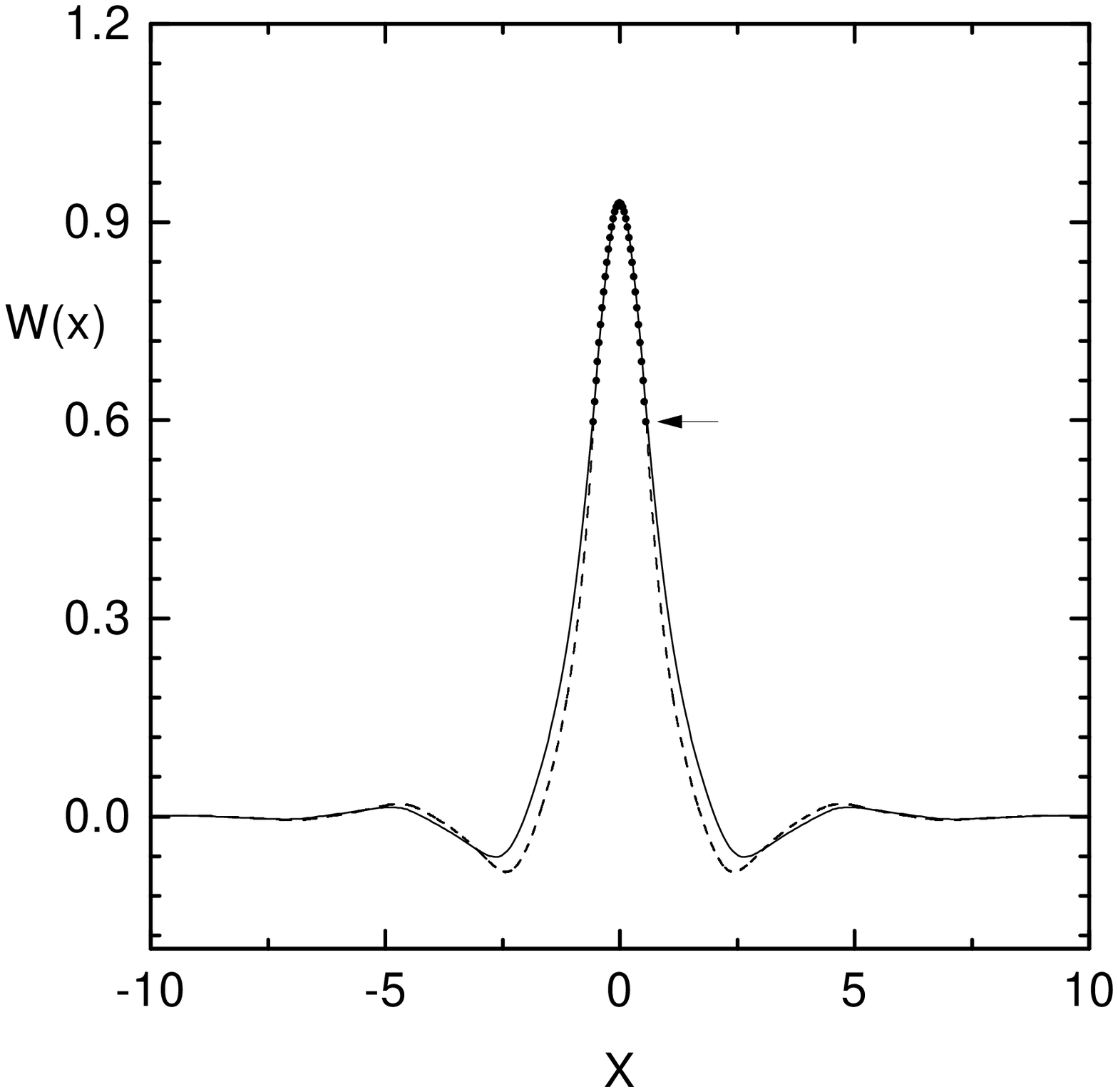}
} \caption{ The Wannier function associated to the lower band of
the one gap potential. The branching points of the elliptic curve
are fixed as in Fig. \ref{fig-1}. The amplitude of the function in
the origin is $W(0)=0.93$. The continuous curve denote the exact
expression obtained from numerical calculations, the dashed line
corresponds to part of the WF approximated by the asymptotic
expansion, while the dotted line denotes the part obtained from
the power expansion near the origin. The arrow shows the point
where the two different analytical expansions are joined.}
\label{fig:3}
\end{figure}
\begin{figure}
\centering
\resizebox{0.65\textwidth}{!}{%
  \includegraphics[clip]{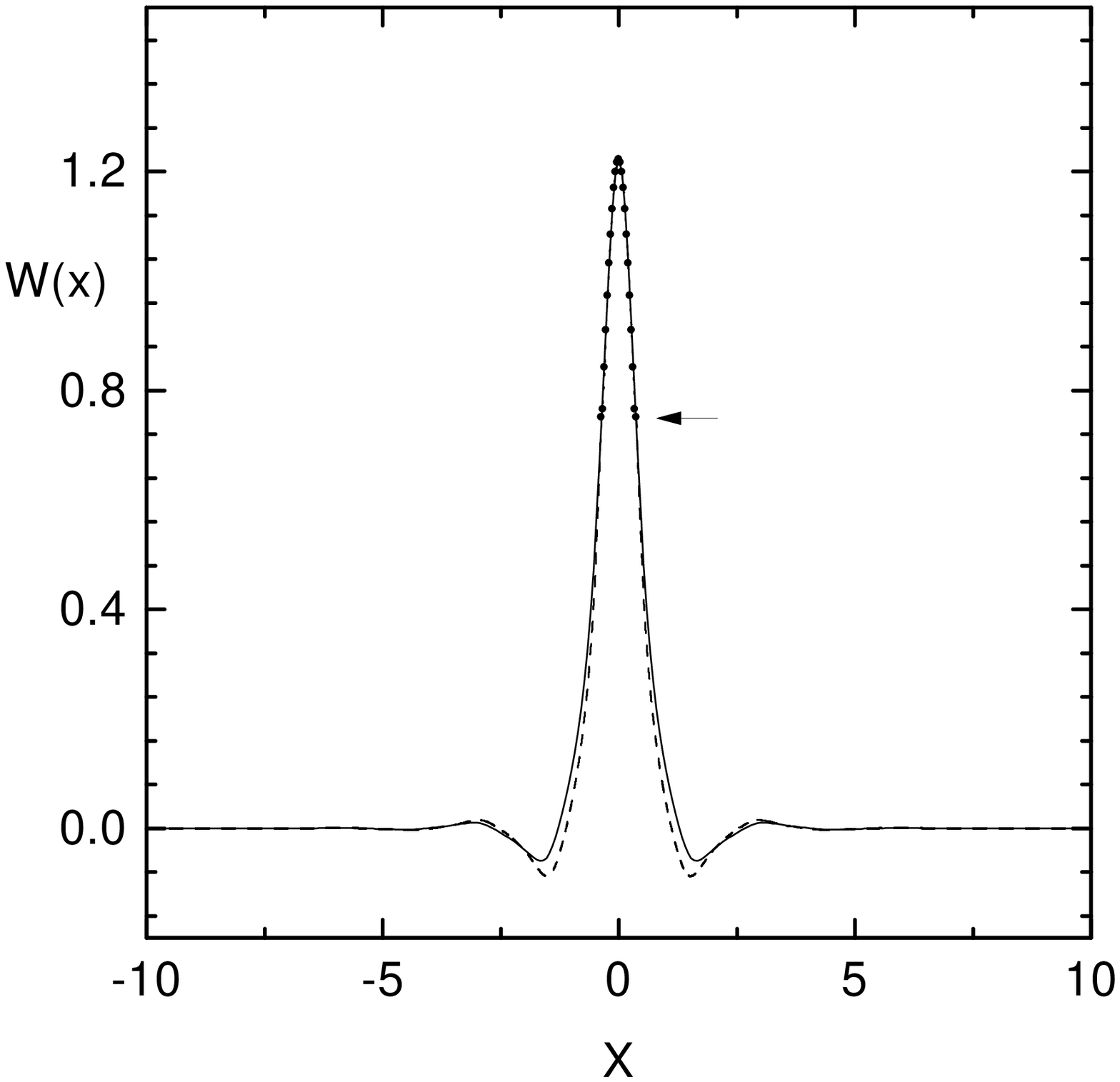}
} \caption{Same as in Fig.\ref{fig:3} but for a different set of
parameters. The branching points of the elliptic curve are
$e_1=6,e_2=-2.0,e_3=-4.0$. The value of the function in the origin
is $W(0)=1.22302$. The continuous curve denote the exact
expression obtained from numerical calculations, the dashed line
corresponds to part of the WF approximated by the asymptotic
expansion, while the dotted line denotes the part obtained from
the power expansion near the origin. The arrow shows the point
where the two different analytical expansions are joined.}
\label{fig:1}
\end{figure}
\begin{figure}
\centering \resizebox{0.65\textwidth}{!}{
  \includegraphics[clip]{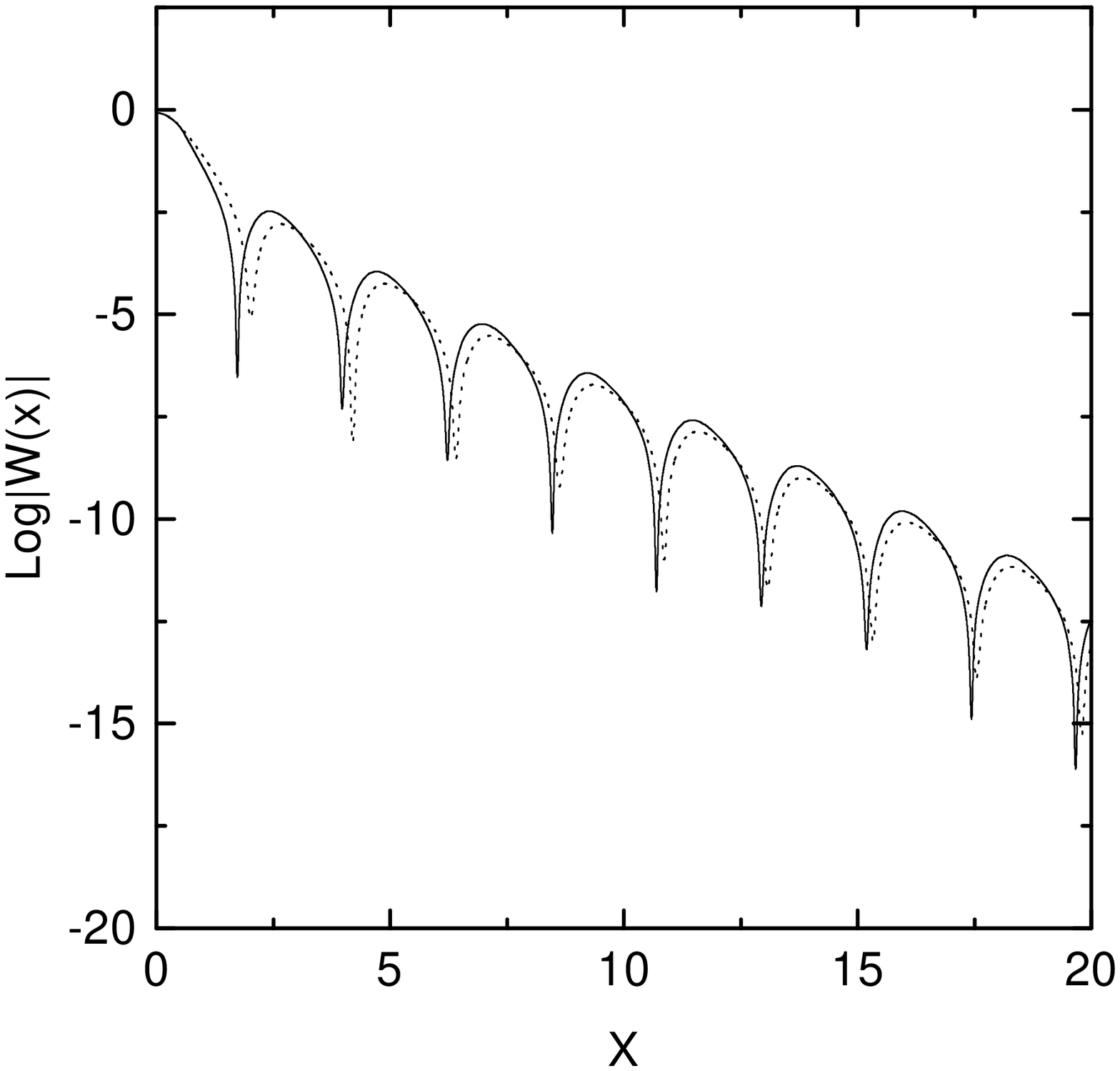}
} \caption{Asymptotic decay of the  Wannier function in
Fig.\ref{fig:3} in semi-log scale. The continuous curve represents
our analytical approximation while the dotted line is obtained
from direct numerical calculations of the WF.} \label{fig:4}
\end{figure}

\begin{figure}
\centering \resizebox{0.65\textwidth}{!}{
   \includegraphics[clip]{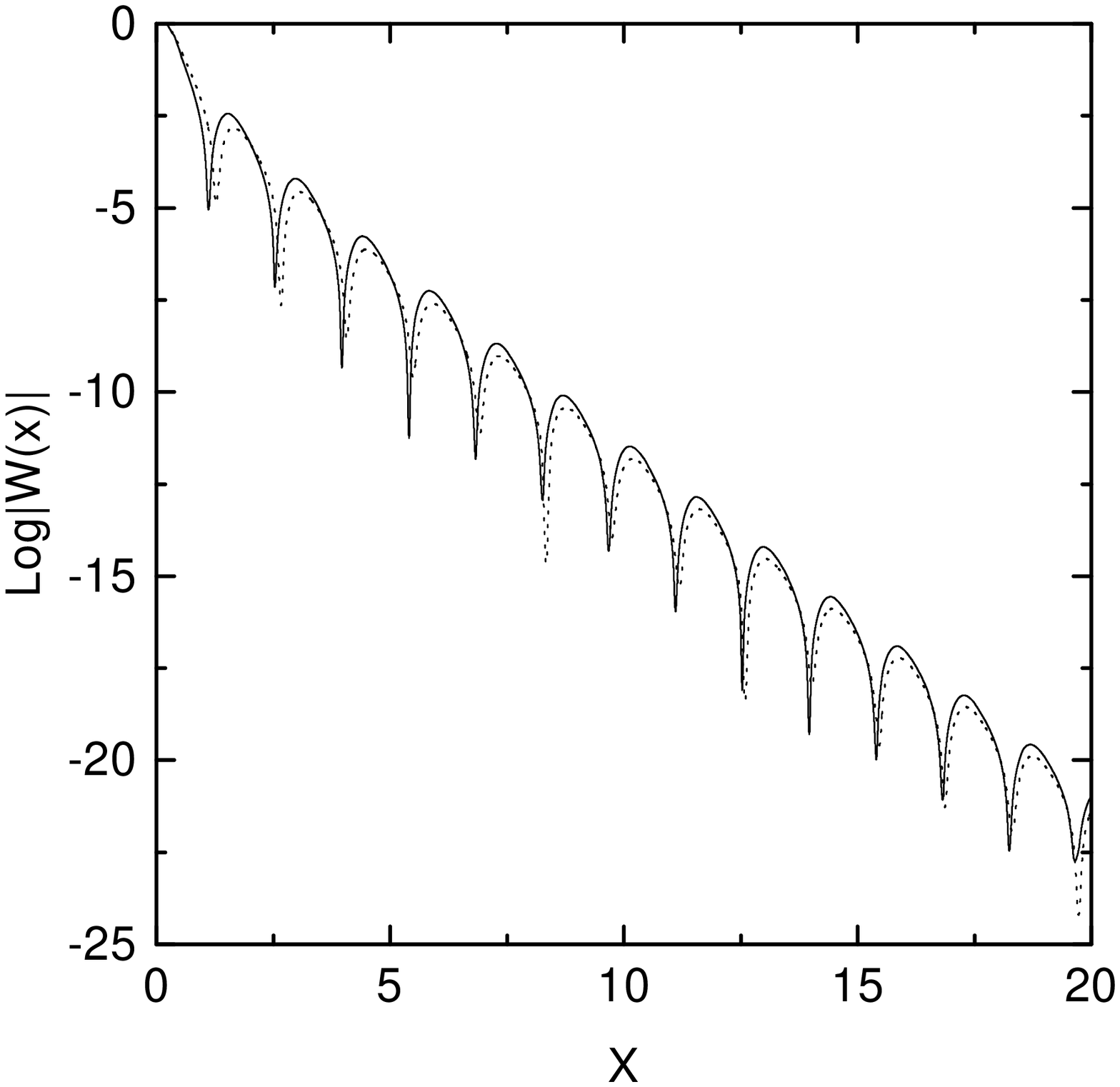}
} \caption{Asymptotic decay of the  Wannier function in Fig.
\ref{fig:1} in semi-log scale. The continuous curve represents our
analytical approximation while the dotted line is obtained from
direct numerical calculations of the WF.} \label{fig:2}
\end{figure}

\section{Approximate analytical expressions of Wannier function and numerical
results} The results of the previous section permit to construct
the following approximate expression of the WF for one-gap
potentials:
\begin{eqnarray}
\label{analiticalapprox}
W(x)&=W_0+W_2x^2+W_4x^4+W_6x^6, \nonumber\\
&\hskip 4cm  \text{for}  \quad |x|\leq x_0 \nonumber \\
&=\mathrm{Re}
\left\{ \frac{\sqrt{-2\imath \omega'}}{\pi}
     \left(
e_1+\frac{\eta^{\prime}}{\omega^{\prime}}\right)^{1/2} \frac{\sigma(v
-u)}{\sigma(v)\sigma(u)}\right.\\
&\left.
\qquad\times \mathrm{exp}\{(u-\omega)\zeta(v)\}
 \left[ \frac{\imath}{2\wp^{\prime}(v)}\right]^{1/4}
\frac{\Gamma\left(\frac34\right)}{x^{\frac34}}\right\}\nonumber \\
&\hskip 4cm \text{for}  \quad |x|>x_0, \nonumber
\end{eqnarray}
where $u=\imath x +\omega$, $v=v_++\omega'$, the coefficients
$W_0\ldots,W_6$ are given by the formulae (\ref{firstw0}) and the
point $x_0$ is chosen so to satisfy the normalization condition
$\| W(x)\|^2=1$.

To check the validity of this  expression we shall compare the WF
obtained from Eq.(\ref{analiticalapprox}) with the one obtained
directly from the definition (\ref{wannier}) by numerical methods,
using the expression of the normalised BF in Proposition
\ref{prop1}. In Fig.\ref{fig-1} we show the band structure
obtained for the one gap potential with parameter values
$e_1=2,e_2=-0.5,e_3=-1.5$, while in Fig. \ref{fig:3} we depict the
WF associated to the lower band. We see that the agreement between
the  analytical approximation and direct numerical calculations is
excellent both in proximity of the origin and far away from it,
these being the regions of validity of the corresponding
expansions. In the intermediate region, however, some discrepancy
appears. It is possible that for some set of  potential parameters
the validity regions of the two expansions (near the origin and
far from the origin) can overlap at some point $x_0$ so that it is
possible to join them into the single smooth analytical
approximation (\ref{analiticalapprox}) which stay close to the
exact numerically curve in the whole spatial domain. Existence of
the point of such a kind is possible only for one-gap potential.
Such a good matching of two expansions is shown in Fig.
\ref{fig:1} where the WF of the lower band for a potential with
another set of parameters is depicted. By comparing Fig.
\ref{fig:1} with Fig. \ref{fig:3} we see that the discrepancy in
the intermediate region is smaller for WF which are more
localised. This can be understood from the fact that a faster
decay of the function (see Figs. \ref{fig:4},\ref{fig:2} below)
allows the asymptotic expansion to work up to points which are
very close to the origin. In Fig.s \ref{fig:4} and \ref{fig:2} we
show the asymptotic decay of the WF depicted in Figs. \ref{fig:3},
\ref{fig:1}, respectively from which we see that a stronger
localization of the function corresponds to a faster asymptotic
decay. The linear decay observed in the semi-log plots of these
figures, is fully consistent with the exponential decay of the WF
of one-gap elliptic potentials predicted by our analysis.

\section{Conclusions}

In this paper we have investigated the properties of the Wannier
functions of the Schr\"odinger operator with one gap potentials.
As a result we have derived the exact value for the amplitude of
the functions in the origin, as well as, an asymptotic expansion
characterising the decay of the function at large distances and a
power series valid in the vicinity of the origin. Using these
expansions  we have constructed approximate analytical expressions
of the Wannier functions and showed that they are in good
agreement with the ones obtained from numerical results.

We remark that the developed approach can be generalised to the case of
finite-gap potentials of more complicated type, like elliptic
finite-gap potentials and general finite-gap potentials.
We shall discuss these problems in a future
publication.

{\bf Acknowledgements}
VZE wish to thank the Dipartimento di Fisica "E.R.Caianiello" of
the University of Salerno for an eight months research grant and
the INFM Unit\'a di Salerno for a three months grant in 2002 year
during which this work was done. EDB acknowledge the Physics
Department of University of Salerno for the hospitality and for
one month financial support in 2002 and 2204 years.
M. S. acknowledges partial support from a
MURST-PRIN-2003 Initiative, and from the European grant LOCNET no.
HPRN-CT-1999-00163.


\begin{thebibliography}{10}
\bibitem{as65}
M.~Abramowitz and I.~A. Stegun.
\newblock {\em Handbook of mathematical functions}.
\newblock Dover, 1965.

\bibitem{bbeim94}
E.~D. Belokolos, A.~I. Bobenko, V.~Z. Enolskii, A.~R. Its, and V.~B. Matveev.
\newblock {\em Algebro {G}eometrical {A}pproach to {N}onlinear {I}ntegrable
  {E}quations}.
\newblock Springer, Berlin, 1994.

\bibitem{bees02}
E.~D. Belokolos, J.~C. Eilbeck, V.~Z. Enolskii, and M.~Salerno.
\newblock Exact energy bands and Fermi surfaces of separable
Abelian potentials.
\newblock {\em J. Phys. A} \textbf{34}, 943--959 (2001)

\bibitem{busch03}
K.Busch, S.Mingaleev, A.Garcia-Martin, M.Schillinger and D.Hermann
\newblock{The Wannier function approach to photonic crystal circuits}
\newblock{\em J.Phys.: Condens. Matter} \textbf{15} (2003) R12333-R1256

\bibitem{ez82} M. Eichler and D. Zagier,
{\em On the zeroes of the {W}eierstrass $\wp$-function}, Math.
Ann., 258, 399-407 (1982)

\bibitem{hevan2001} L.~He and D.~Vanderblit,
\newblock{\em Exponential {D}ecay {P}roperties of {W}annier
{F}unctions and {R}elated {Q}uantities  }
\newblock {\em Phys. Rev. Lett.} \textbf{86}, no. 23, 5341-5344 (2001)


\bibitem{cloize64} J.~Cloizeaux,
\newblock{\em Analytical Properties of $n$-Dimensional Energy Bands and
            Wannier Functions }
\newblock {\em Phys. Rev.} \textbf{135}, no. 3A, A698-A707 (1964)

\bibitem{cloize64a} J.~Cloizeaux,
\newblock{\em Energy Band and Projection Operators in a Crystal:
 Analytic and asymptotic Properties }
\newblock {\em Phys. Rev.} \textbf{133}, no. 3A, A685-A697 (1964)


\bibitem{fed87}
M.~Fedoriuk.
\newblock {\em Asymptotics: {I}ntegrals and {S}eries (in Russian)}.
\newblock Nauka, Moscow (1987).


\bibitem{gh03}
F.~Gesztesy and H.~Holden.
\newblock {\em {S}oliton {E}quations and {T}heir {A}lgebro-{G}eometric
  {S}olutions. $(1+1)$-{D}imensional {C}ontinuous {M}odels}.
\newblock Cambridge University Press, Cambridge, U.K., 2003.

\bibitem{gkkm98}
M.~Gl\"uck, A.R.~Kolowsky, H.J.~Korsch and N.~Moiseyev
\newblock{Calculation of Wannier-Bloch and Wannier-Stark states}
\newblock{\em Eur.Phys.J.D.} \textbf{4}, 239-246 (1998)

\bibitem{im75}
A.~R. Its and V.~B. Matveev.
\newblock Schr\"odinger
operators with a finite-band spectrum  and the $N$-soliton
  solutions of the {K}orteweg-de {V}ries equation.
\newblock {\em Teoret. Mat. Fiz.} \textbf{23}, 51--68 (1975)

\bibitem{kohn59}
W.~Kohn.
\newblock Analytic {P}roperties of {B}loch {W}aves and {W}annier {F}unction.
\newblock {\em Phys. Rev.} \textbf{115}, 809--821 (1959)

\bibitem{kohn73}
W.~Kohn.
\newblock Construction of {W}annier {F}unctions and Applications to
Energy Band.
\newblock {\em Phys. Rev. B} \textbf{7}, no. 10, 4388-4398 (1973)

\bibitem{kr77}
I.~M. Krichever.
\newblock The method of algebraic geometry in the theory of nonlinear
  equations.
\newblock {\em Russian. Math. Surveys} \textbf{32}, 180--208 (1977)

\bibitem{wan37}
G.~Wannier.
\newblock The {S}tructure of {Electronic} {E}xcitation {L}evels
 in {I}nsulating {C}rystals.
\newblock {\em Phys. Rev.} \textbf{52}, 191--197 (1937)

\bibitem{wan60}
G.~Wannier.
\newblock Wave Functions and Effective Hamiltonian
for Bloch Electrons in an Electric Field
\newblock {\em Phys. Rev.} \textbf{117}, no. 2  432-439 (1937)

\bibitem{wilk98}
M.~Wilkinson.
\newblock Wannier Functions for Lattices in a
Magnetic Field
\newblock {\em J. Phys.: Condens. Matter} \textbf{10},7407-7427 (1998)


\bibitem{zmnp80} V.~E. Zakharov, S.~V. Manakov,
S.~P. Novikov and L.~P. Pitaevski.
\newblock{\em Soliton theory: inverse scattering method (in Russian)},
\newblock Nauka, Moscow, (1980)



\end{thebibliography}
\end{document}